\documentclass[12pt,epsfig,a4]{article} 
\newcommand{\vs}{\vspace{-0.25cm}}
\usepackage{amsmath,graphicx}
\begin{document} 
\begin{center}
{\Large{\bf Skyrme interaction to second order in nuclear 
matter}\footnote{This work 
has been supported in part by DFG and NSFC (CRC110).}  }  

\medskip

 N. Kaiser \\
\medskip
{\small Physik-Department T39, Technische Universit\"{a}t M\"{u}nchen,
   D-85747 Garching, Germany}
\end{center}
\medskip
\begin{abstract}
Based on the phenomenological Skyrme interaction various density-dependent 
nuclear matter quantities are calculated up to second order in many-body 
perturbation theory. The spin-orbit term as well as two tensor terms 
contribute at second order to the energy per particle. The simultaneous 
calculation of the isotropic Fermi-liquid parameters provides a rigorous check 
through the validity of the Landau relations. It is found that published 
results for these second order contributions are incorrect in most cases. In 
particular, interference terms between $s$-wave and $p$-wave components of the 
interaction can contribute  only to (isospin or spin) asymmetry energies. 
Even with nine adjustable parameters, one does not obtain a good description 
of the empirical nuclear matter saturation curve in the low density region 
$0<\rho<2\rho_0$. The reason for this feature is the too strong 
density-dependence $\rho^{8/3}$ of several second-order contributions. The 
inclusion of the density-dependent term ${1\over 6}t_3 \rho^{1/6}$ is 
therefore indispensable for a realistic description of nuclear matter in the 
Skyrme framework.
    
\end{abstract}

\section{Introduction and summary}
Mean-field theories are widely used to analyze a large variety of many-body 
systems for which the independent particle picture can be applied. In nuclear 
physics, mean-field approaches generally lead to satisfactory results for bulk 
properties of atomic nuclei, such as their masses, radii and ground-state 
deformations. However, in some cases improvements beyond the mean-field 
treatment are definitely needed. For instance, mean-field models do not 
predict accurately the single-particle spectra (i.e. energies, spectroscopic 
factors and fragmentations of single-particle states), especially close to 
the Fermi energy. A possibility for improvement is given by introducing the 
coupling between individual nucleon degrees of freedom and collective degrees 
of freedom.

Phenomenological Skyrme forces have been  used widely and successfully for 
non-relativistic nuclear structure calculations of medium-mass and heavy 
nuclei \cite{bender,stone}. In effect such purely phenomenological two-body 
interactions can be viewed as to provide a convenient parameterization of the 
nuclear energy density functional on which the self-consistent mean-field 
treatment (in the form of solving the Kohn-Sham equations) can be based. In 
order to go beyond the common mean-field (Hartree-Fock) approximation, 
second-order calculations with the phenomenological Skyrme force have been 
pursued in recent works by Moghrabi, Grasso et al.\,\cite{moghrabi1,moghrabi2,
moghrabi3,moghrabi4}. These exploratory studies have been concerned with the 
second-order contributions to the equation of state of infinite 
isospin-symmetric nuclear matter and pure neutron matter. Since the 
(two-body) Skyrme interaction is of zero range, divergences occur at second 
order in many-body perturbation theory, which have been treated either by 
cutoff regularization \cite{moghrabi1,moghrabi2} or by dimensional 
regularization \cite{moghrabi3}. It has been concluded in 
refs.\,\cite{moghrabi2,moghrabi3} that after readjusting the parameters of 
the Skyrme interaction a simultaneous good reproduction of the equation of 
state of isospin-symmetric and isospin-asymmetric nuclear matter can be 
achieved at second order. 

However, a closer inspection of the published expressions for the 
second-order contributions in ref.\,\cite{moghrabi3} raises doubts about 
their completeness and correctness. The first deficit is the omission of a 
second-order contribution from the spin-orbit term proportional to $W_0$. 
This $p$-wave interaction term averages to zero in infinite matter at first 
order, but not at second order (see e.g. appendix A in ref.\,\cite{spinorbit}
for the second-order contributions from a ``boson-exchange'' spin-orbit 
interaction). Secondly, the stated expressions 
\cite{moghrabi3} for the energy per particle of isospin-symmetric nuclear 
matter and pure neutron matter include interference terms between $s$-wave 
and $p$-wave components of the Skyrme interaction (i.e. terms  
proportional to $t_0t_2$ and $t_1t_2$). It is very questionable how such a 
mixing of different parities could be induced by the (homogeneous) nuclear 
medium. Moreover, the expressions for the cutoff-independent second-order 
contributions given in a later paper \cite{moghrabi4} by the same authors 
disagree partly with those published in ref.\,\cite{moghrabi3}.

These apparent inconsistencies provide good reason to reconsider and extend the 
calculation of the Skyrme interaction to second order in nuclear matter. The 
present paper is organized as follows: In section 2, the Skyrme force is 
introduced as a Galilei-invariant two-nucleon contact-interaction in momentum 
space. On this basis, first-order Hartree-Fock results are rederived for a 
variety of density-dependent nuclear matter quantities, such as the energy 
per particle and the (isospin and spin) asymmetry energy.  Section 3 is 
devoted to the second-order calculation in many-body perturbation theory. We 
employ a scheme where the in-medium (particle-hole) propagator is decomposed 
into the vacuum propagator and a medium-insertion. This scheme has the 
advantage that ultraviolet divergences appear only in the second-order 
Hartree and Fock diagrams with two medium-insertions. The pertinent one-loop 
rescattering in vacuum is treated by dimensional regularization and shown to 
vanish identically. As a consequence the genuine second-order terms arise 
entirely from the Hartree and Fock diagrams with three medium-insertions. The 
occurring finite integrals over three Fermi spheres can be 
solved analytically and the results for the nuclear matter quantities consist 
of even powers of the Fermi momentum $k_f$ multiplied with products of the 
Skyrme parameters and a numerical coefficient of the form $\ln2+r_j$, where 
$r_j$ is some rational number. It is found that the energy per particle of 
isospin-symmetric nuclear matter $\bar E(k_f)$ and pure neutron matter 
$\bar E_n(k_n)$ do not include any $s$-wave and $p$-wave interference terms at 
second order. However, such $sp$-interference terms show up in the (isospin 
and spin) asymmetry energies. The reason and condition for $sp$-mixing to 
emerge is explained in appendix A by considering the in-medium loop for a 
proton-neutron pair in nuclear matter with unequal proton and neutron 
densities. In section 4, the effective nucleon mass and the isotropic 
Fermi-liquid parameters are calculated up to second order. These results allow 
for a rigorous check through the validity of several Landau relations 
\cite{landau}. In section 5, the effects of tensor interactions at second 
order are computed. Actually, with the inclusion of two independent tensor 
terms the Skyrme force gets completed to the most general two-nucleon 
contact-interaction quadratic in momenta. In section 6, the iteration of the 
Skyrme force with the one-photon exchange is considered. This mechanism 
generates a charge-symmetry breaking contribution to the energy density of 
protons, which could help to resolve the Nolen-Schiffer 
anomaly \cite{nolen}. This anomaly refers to the observation that the 
binding energy differences of mirror nuclei cannot be explained by the Coulomb 
interaction of the protons alone. Remarkably, one finds an infrared-finite 
result for the energy per proton $\bar E_p(k_p)$ even though a regulator 
photon mass $m_\gamma$ has be introduced initially. Section 7 deals 
with some provisional parameter fits. It is found that with nine adjustable 
parameters in the complete second-order expression, one does not obtain a 
good reproduction of a realistic nuclear matter saturation curve in the 
low-density region $0<\rho<2\rho_0$. The reason for this failure lies in the 
too strong density-dependence $\rho^{8/3}$ of several second-order 
contributions. From this observation one can conclude that the inclusion of 
(ad hoc) density-dependent terms is indispensable for a realistic description 
of nuclear matter in the Skyrme approach. At a formal level, these are 
taken into account by performing appropriate parameter substitutions, 
$t_0\to t_0+{1\over 6}t_3\rho^{1/6}$ and  $t_0x_0\to t_0x_0 +{1\over 6}t_3x_3
\rho^{1/6} $, in all first- and second-order expressions derived in this work.  

Let us clarify that the main purpose of this paper is to compile the 
(correct) analytical expressions for various nuclear matter quantities as they 
arise from a second-order calculation with a two-body contact-interaction of 
the Skyrme type. The question of whether improvements are possible in 
comparison to a first-order mean-field approach has to be explored in future 
studies.     
      
\section{First-order Hartree-Fock calculation}
This section serves mostly for orientation. We introduce various 
density-dependent nuclear matter quantities and specify their first-order 
Hartree-Fock contributions. The Skyrme interaction is a zero-range two-nucleon 
contact-interaction with transition matrix-element in momentum-space: 
\begin{equation} V_{\rm Sk} = t_0(1+x_0 P_\sigma) +{t_1 \over 2}(1+x_1 P_\sigma) 
(\vec q_{\rm out}^{\,2}+ \vec q_{\rm in}^{\,2})+t_2(1+x_2 P_\sigma)\, 
\vec q_{\rm out}\cdot \vec q_{\rm in} + i\, W_0(\vec \sigma_1+\vec \sigma_2)
\cdot (\vec q_{\rm out}\times \vec q_{\rm in}) \,.\end{equation}
Here, $P_\sigma\!=\!(1+\vec \sigma_1\cdot\vec \sigma_2)/2$ is the 
spin-exchange operator and  $\vec q_{\rm in} = (\vec p_1-\vec p_2)/2$ and  
$\vec q_{\rm out} = ({\vec p_1}\,\!\!'-\vec p_2\,\!\!')/2$ denote half of the 
momentum difference in the initial and final state, respectively. This 
particular form of $V_{\rm Sk}$, involving only external momentum differences, 
is dictated by Galilei invariance (and invariance under time-reversal: 
$\vec q_{\rm in} \leftrightarrow - \vec q_{\rm out}$, $\vec \sigma_{1,2}  \to 
- \vec \sigma_{1,2}$). The individual terms in eq.(1) stand for: 
$s$-wave contact interaction ($t_0$), $s$-wave effective range correction 
($t_1$), $p$-wave  contact interaction ($t_2$), and spin-orbit interaction 
($W_0$). The general contact-potential up to order $p^2$, as it is commonly 
used in chiral effective field theory \cite{evgeni,hammer,machleidt}, has two 
additional tensor terms which will be studied separately in section 5. Note 
that we have not explicitly written in eq.(1) the density-dependent Skyrme 
term ${1\over 6}t_3(1+x_3 P_\sigma)\rho^{1/6}$. Formally, it can be included 
through the substitution of parameters $t_0\to t_0 +{1\over 6}t_3\rho^{1/6} $ 
and $t_0x_0\to t_0x_0+{1\over 6}t_3x_3\rho^{1/6}$ in all following formulas. 
However, from a rigorous point of view such density-dependent couplings will 
lead to a conflict with basic many-body relations (Landau relations 
\cite{landau} and Hugenholtz-Van-Howe theorem \cite{hvh}), whenever 
derivatives with respect to density $\rho$ are involved.

\begin{figure}[h!]
\begin{center}
\includegraphics[scale=0.53,clip]{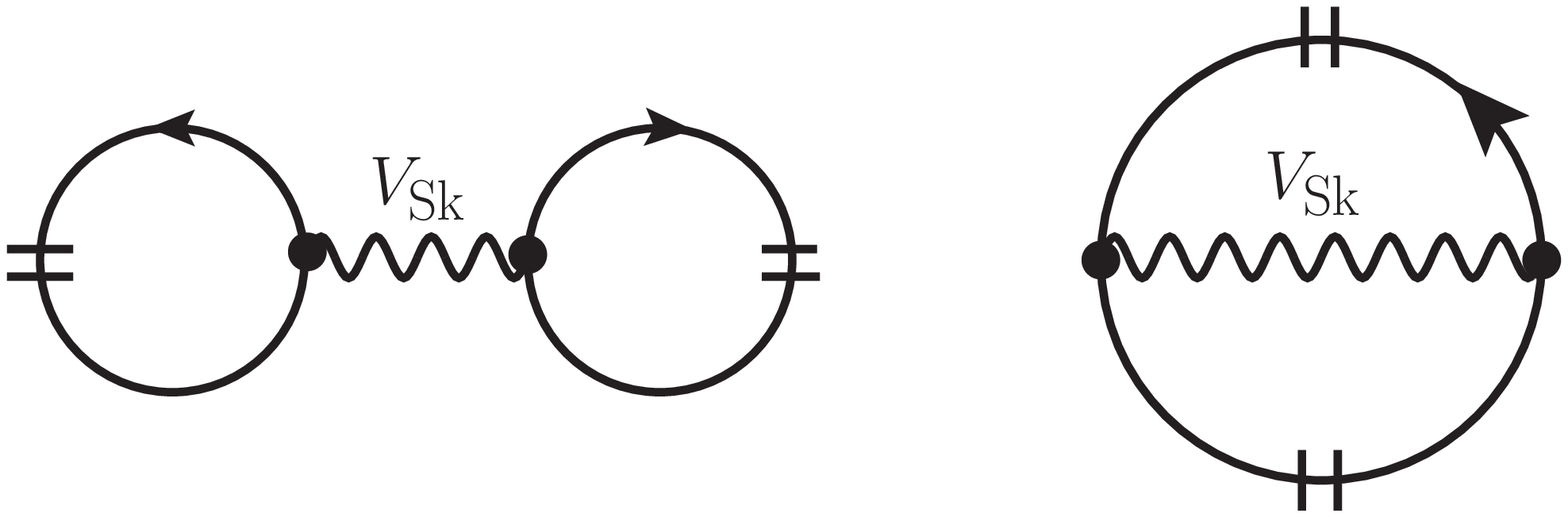}
\includegraphics[scale=0.57,clip]{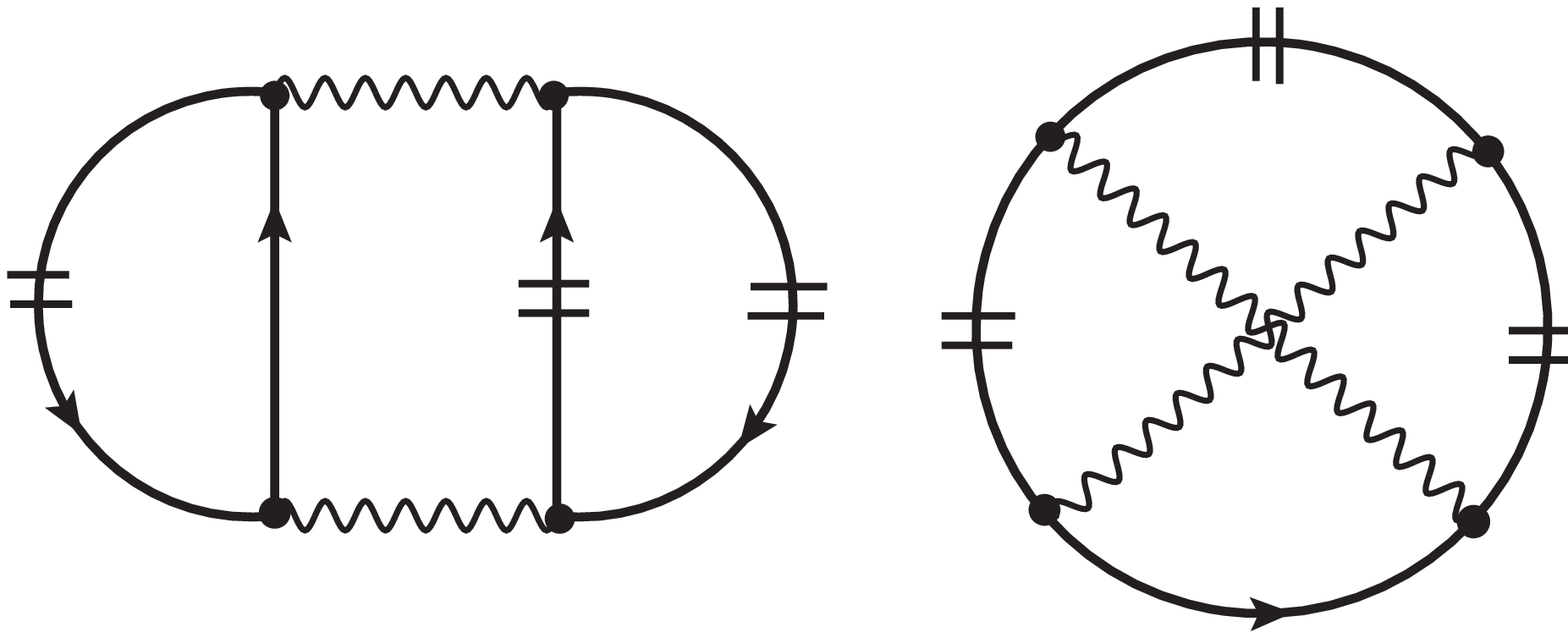}
\end{center}
\vspace{-.6cm}
\caption{Upper row: First-order Hartree and Fock diagrams with two 
medium-insertions and combinatoric factor $1/2$. The wavy line represents the 
Skyrme interaction $V_{\rm Sk}$. Lower row: Second-order Hartree and Fock 
diagrams with three medium-insertions and combinatoric factor $1$.}
\end{figure}

The first-order Hartree and Fock diagrams generated by $V_{\rm Sk}$ are shown 
in the upper part of Fig.\,1. The short double-line symbolizes the 
medium-insertion (i.e. the difference between the in-medium and vacuum 
nucleon-propagator) proportional to the step-function $\theta(k_f-|\vec p_j|)$.
The Fermi momentum $k_f$ is related the (total) nucleon density in the usual 
way, $\rho=2k_f^3/3\pi^2$. Performing the appropriate spin- and isospin-traces 
and integrals over two Fermi spheres, one obtains for the energy per particle 
of isospin-symmetric nuclear matter the well-known first-order expression:
\begin{equation} \bar E(k_f)^{(\rm 1st)} = {k_f^3 \over 4\pi^2} \bigg\{ t_0 
+{k_f^2 \over 10}\big[3t_1+t_2(5+4x_2)\big]\bigg\} \,.\end{equation}
In the same way, but leaving out the isospin-trace, the energy per particle 
of pure neutron matter comes out as:
\begin{equation} \bar E_n(k_n)^{(\rm 1st)} = {k_n^3 \over 4\pi^2} \bigg\{ 
{t_0 \over 3}(1-x_0)+{k_n^2 \over 10}\big[t_1(1-x_1)+3t_2(1+x_2)\big]\bigg\} 
\,,\end{equation}
with $k_n$ the neutron Fermi momentum related to the neutron density by 
$\rho_n= k_n^3/3\pi^2$. Another important quantity of high interest is the 
isospin asymmetry energy of nuclear matter. An isospin-asymmetric configuration 
with unequal proton and neutron Fermi momenta $ k_{p,n}=k_f(1\mp \delta)^{1/3}$ 
is realized by the substitution: 
\begin{equation}\theta(k_f-|\vec p_j|) \to { 1+\tau_3 \over 2}\,\theta(k_p
-|\vec p_j|)+ {1-\tau_3 \over 2}\,\theta(k_n-|\vec p_j|)\,,\end{equation}
for the medium-insertion and an additional expansion of the (asymmetric)
energy per particle:
\begin{equation} \bar E_{\rm as}(k_p,k_n)= \bar E(k_f)+\delta^2 
A(k_f)+{\cal O}(\delta^4)  \,,\end{equation}
around $\delta =0$ defines the density-dependent isospin asymmetry $A(k_f)$ 
of nuclear matter. The first-order Hartree-Fock contribution from the Skyrme 
interaction $V_{\rm Sk}$ has the well-known form:
\begin{equation} A(k_f)^{(\rm 1st)} = {k_f^3 \over 12\pi^2} \bigg\{ -t_0
(1+2x_0)+{k_f^2\over 3}\big[t_2(4+5x_2)-3t_1 x_1\big]\bigg\}\,.\end{equation}
In an analogous way the density-dependent spin asymmetry energy $S(k_f)$ of 
nuclear matter can be constructed. The substitution in spin-space:
\begin{equation}\theta(k_f -|\vec p_j|)\to {1+\sigma_3\over 2}\,\theta(k_\uparrow
-|\vec p_j|)+{1-\sigma_3\over 2}\,\theta(k_\downarrow-|\vec p_j|)\,,\end{equation}
with $ k_{\uparrow,\downarrow}=k_f(1\pm \eta)^{1/3}$ the Fermi momenta of the 
spin-up and spin-down nucleons, is followed by an expansion of the energy per 
particle  of spin-polarized nuclear matter around $\eta =0$:
\begin{equation} \bar E_{\rm pol}(k_\uparrow,k_\downarrow)= \bar E(k_f)
+\eta^2 S(k_f)+{\cal O}(\eta^4)  \,.\end{equation}
In the absence of any quantitative empirical knowledge, the minimal condition 
to be satisfied by the spin asymmetry energy is that it must be positive: 
$S(k_f)>0$. This guarantees the spin-stability of nuclear matter by excluding 
a phase transition to a ferromagnetic state of lower energy density. The 
first-order Hartree-Fock contribution to $S(k_f)$ from the Skyrme interaction 
$V_{\rm Sk}$ reads: 
\begin{equation} S(k_f)^{(\rm 1st)} = {k_f^3 \over 12\pi^2} \bigg\{ t_0 
(2x_0-1)+{k_f^2 \over 3}\big[3t_1 x_1+t_2(4+5x_2)\big]\bigg\} \,.\end{equation}
One notices that the expression for $S(k_f)^{(\rm 1st)}$ can be obtained from 
$A(k_f)^{(\rm 1st)}$ in eq.(6) by just changing the signs of the two parameters 
$x_0$ and $x_1$. The reason behind this connection is an equi- valent form of 
the contact-interaction which results from $V_{\rm Sk}$ in eq.(1) through the 
substitutions: 
\begin{eqnarray} t_0(1+x_0 P_\sigma) & \to &   t_0(1-x_0 P_\tau)\,,\nonumber \\
 t_1(1+x_1 P_\sigma) & \to &   t_1(1-x_1 P_\tau)\,,  \nonumber \\
 t_2(1+x_2 P_\sigma) & \to &   t_2(1+x_2 P_\tau)\,, \end{eqnarray}
with $P_\tau= (1+\vec \tau_1\!\cdot\!\vec \tau_2)/2$ the isospin-exchange 
operator. The known eigenvalues of $P_\sigma$ and $P_\tau$ are $(-1)^{S+1}$ and  
$(-1)^{I+1}$ with $S,I \in \{0,1\}$ the total spin and isospin quantum numbers. 
Since the parameters $t_0$ and $t_1$ act only in $s$-wave states with $S+I=1$, 
one has effectively for $s$-waves the equality $P_\sigma=-P_\tau$. Likewise, the 
parameter $t_2$ acts only in $p$-wave states with $S+I=0,2$ and therefore 
for $p$-waves the opposite equality $P_\sigma = P_\tau$ holds. Another obvious 
consequence of the equivalent interaction is that only the combinations 
$t_0(1-x_0)$,  $t_1(1-x_1)$ and $t_2(1+x_2)$ can contribute in neutron matter, 
since $P_\tau=1$ for a two-neutron state.  This is exemplified by calculating 
the first-order contribution to the spin asymmetry energy of pure neutron 
matter:
\begin{equation} S_n(k_n)^{(\rm 1st)} = {k_n^3 \over 12\pi^2} \bigg\{t_0(x_0-1) 
+ {k_n^2\over 3} \big[t_1(x_1-1)+3t_2(1+x_2) \big]\bigg\} \,.\end{equation}
Again, the density-dependent quantity $S_n(k_n)$ must be positive: 
$S_n(k_n)>0$, in order to prevent neutron matter from a ferromagnetic 
instability at low and moderate densities. We conclude this section by giving
the contributions from the relativistically improved kinetic energy:  
\begin{equation} \bar E(k_f)^{(\rm kin)} ={3k_f^2 \over 10M}-{3k_f^4 
\over 56 M^3} \,, \qquad\quad \bar E_n(k_n)^{(\rm kin)} ={3k_n^2 \over 10M}
-{3k_n^4 \over 56 M^3} \,,\end{equation}
\begin{equation} A(k_f)^{(\rm kin)} = S(k_f)^{(\rm kin)} ={k_f^2 \over 6M}
-{k_f^4 \over 12 M^3} \,,\qquad\quad S_n(k_n)^{(\rm kin)} ={k_n^2 \over 6M}
-{k_n^4 \over 12 M^3} \,,\end{equation}
where $M=939\,$MeV denotes the average nucleon mass.
\section{Second-order Hartree-Fock calculation}
In this section we derive analytical expressions for the contributions to 
the introduced nuclear matter quantities as they arise from a second-order 
calculation with the Skyrme interaction $V_{\rm Sk}$. At second order in 
many-body perturbation theory the energy density is represented by closed 
three-loop Hartree and Fock diagrams with either two or three medium-insertions.
The first set involves the rescattering of two nucleons in the vacuum (i.e. 
without Pauli blocking corrections). For the contact-interaction $V_{\rm Sk}$ 
this one-loop diagram leads to principal-value integrals of the form:
\begin{equation} -\hspace{-0.45cm}\int\!{d^3 l\over(2\pi)^3}\,{M\over\vec 
l^{\,2}-\vec q^{\,2}}\Big\{1,\vec l^{\,2}, (\vec l^{\,2})^2 , (\vec l \!\cdot\! 
\vec q\,)^2 \Big\} \,,\end{equation}
with $\vec q = (\vec p_1-\vec p_2)/2$, where $\vec p_{1,2}$ are two external 
momenta inside the Fermi sphere $|\vec p_{1,2}|<k_f$. The first factor in 
eq.(14) is the energy denominator and the other factors in curly brackets 
come from the Galilei-invariant contact-interaction. After angular 
integration and partial fraction decomposition of the radial integrand one 
gets power divergences of the form $\int_0^\infty\!dl\{1,l^2,l^4\}$, which are 
set to zero in dimensional regularization. The remaining convergent 
principal-value integral  $-\hspace{-0.38cm}\int_0^\infty\!dl(l^2-q^2)^{-1}=0$ 
vanishes identically. As a side remark we note that cutoff regularization 
would provide additional terms $\Lambda^{1,3,5}$, $\Lambda^{1,3} \vec q^{\,2}$ 
and $\Lambda (\vec q^{\,2})^2$ proportional to odd powers of the cutoff 
$\Lambda$. For the first two the $\Lambda$-dependence could be absorbed into 
the coupling constants $t_j$ and $t_jx_j$, while the last one would require 
the introduction of a four-derivative contact-interaction.

In dimensional regularization (or after renormalization) one is therefore 
left with the Hartree and Fock diagrams with three medium-insertions shown in 
the lower part of Fig.\,1. These lead to (finite) principal-value integrals 
over three Fermi spheres of the form:
\begin{equation} -\hspace{-0.745cm}\int\limits_{|\vec p_j|<k_f}\!\!\!\!\!{d^3p_1 
d^3p_2 d^3p_3 \over (2\pi)^9}\,{M \over \vec q_1\!\cdot\! \vec q_2}\, 
H(\vec q_1^{\,2}, \vec q_1\!\cdot\!\vec q_2,\vec q_2^{\,2})  \,,\end{equation}
with $\vec q_{1,2} =\vec p_3-\vec p_{1,2}$. The first factor $M/(\vec q_1\!\cdot
\! \vec q_2$) is again the energy denominator and the function $H(...)$ is a 
polynomial of total degree two or less in its three variables. In order to  
evaluate eq.(15), one eliminates $\vec p_2$ in favor of $\vec q_2$ and 
integrates over a shifted Fermi sphere $|\vec q_2-\vec p_3|<k_f$. The upper 
boundary for the $dq_2$-integration is then $p_3 y+\sqrt{k_f^2-p_3^2+p_3^2 y^2}$, 
with the directional cosine $y = \hat q_2\!\cdot\! \hat p_3$. By applying this 
procedure to evaluate the nine-dimensional principal-value integrals, one is 
able to obtain analytical results for all second-order contributions. Let us 
also note that for $s$-wave interactions $\sim t_0,t_1$ the Hartree and Fock 
diagram contribute with opposite sign, whereas for $p$-wave interactions 
$\sim t_2,W_0$ the Hartree and Fock diagram contribute with equal sign. 

Putting all pieces together one ends up with the following
expression for the second-order contribution from $V_{\rm Sk}$ to the 
energy per particle of isospin symmetric nuclear matter:
\begin{eqnarray} \bar E(k_f)^{(\rm 2nd)} &=&{M k_f^4 \over 280 \pi^4} \bigg\{ 
3t_0^2(1+x_0^2)(11-2\ln 2) +{2k_f^2 \over 9}t_0 t_1(1+x_0 x_1)
(167-24\ln 2)\nonumber \\ && +{k_f^4 \over 396} t_1^2(1+x_1^2)(4943-564\ln 2)
+{k_f^4 \over 1188}t_2^2(5+8x_2+5 x_2^2)(1033-156\ln 2)\nonumber \\ && 
+{4k_f^4 \over 99}W_0^2 (631-102\ln 2) \bigg\}\,.\end{eqnarray} 
The first important observation one can make is that there are no interference 
terms between  $s$-wave and $p$-wave components of the Skyrme interaction 
$V_{\rm Sk}$. A detailed comparison to the expressions written in eqs.(16,17) 
of ref.\,\cite{moghrabi3} reveals that therein only the term proportional 
to $t_0^2(1+x_0^2)$ is correct. The different form $\bar E(k_f)^{(\rm 2nd)}=3M k_f^4
(A_0+A_3 k_f^2+A_5 k_f^4)/2\pi^4$ given in eq.(5) of the later paper 
\cite{moghrabi4} by the same authors represents no improvement either. It be 
should noted that ref.\,\cite{moghrabi3} did not use the advantageous scheme 
of expanding in the number of medium-insertions and therefore divergent 
integrals of the form eq.(14) as well as finite integrals of the form eq.(15) 
had to be continued together to $d\ne 3$ space-dimensions. Moreover, the 
Galilei-invariant formulation of the interaction $V_{\rm Sk}$ in momentum-space 
is important. 

By adding half of the (previous) Hartree contribution to the Fock 
contribution one arrives at the energy per particle of pure neutron matter:
\begin{eqnarray} \bar E_n(k_n)^{(\rm 2nd)} &=&{M k_n^4 \over 280\pi^4}\bigg\{ 
t_0^2(1-x_0)^2(11-2\ln 2) +{2k_n^2 \over 27}t_0 t_1(1-x_0)(1-x_1)
(167-24\ln 2)\nonumber \\ && +{k_n^4 \over 1188} t_1^2(1-x_1)^2(4943-564\ln 2)
+{k_n^4 \over 396}t_2^2(1+x_2)^2(1033-156\ln 2) \nonumber \\ && 
+{8k_n^4 \over 297}W_0^2 (631-102\ln 2) \bigg\}\,.\end{eqnarray}
As expected from the consideration of the equivalent interaction in 
section 2, the second-order result $\bar E_n(k_n)^{(\rm 2nd)}$ involves only the
parameter combinations $t_0(1-x_0),\, t_1(1-x_1)$ and $t_2(1+x_2)$. 
The corresponding expression given in ref.\,\cite{moghrabi3} is correct only 
with respect to the term proportional to $t_0t_1(1-x_0)(1-x_1)$, and the 
(simplest) term proportional $t_0^2(1-x_0)^2$ misses a factor $2$ therein.
Note that the factor $(1033-156\ln 2)/385$ appears also in eq.(32) of 
ref.\cite{pwave}, where the second-order contribution to $\bar E_n(k_n)$ 
has been extracted from the resummation to all orders of particle-hole ladder 
diagrams generated by a spin-independent $p$-wave contact interaction.

The second-order contribution to the isospin asymmetry energy $A(k_f)$ is 
more difficult to calculate. One has to evaluate now second derivatives 
with respect to $\delta$ at $\delta =0$ of principal-value integrals over 
three Fermi spheres with different radii $k_{p,n} = k_f(1\mp \delta)^{1/3}$.   
The complete expression for the second-order contribution from the Skyrme 
interaction $V_{\rm Sk}$ to the isospin asymmetry energy of nuclear matter 
reads: 
{\allowdisplaybreaks   
\begin{eqnarray} A(k_f)^{(\rm 2nd)} &=&{M k_f^4 \over 60\pi^4}\bigg\{ 
t_0^2\big[(1+x_0^2)(1-2\ln 2) -10x_0\big] \nonumber \\ && + {k_f^2 \over 42}
t_0 t_1\big[4(1+x_0x_1)(51-22\ln 2) +7(x_0+x_1)(4\ln 2-57)\big]\nonumber \\ && 
+{k_f^4 \over 63}t_1^2\Big[(1+x_1^2)\Big({1301 \over 6}-46\ln 2\Big) 
+x_1(48\ln 2-607)\Big]\nonumber \\ && +{k_f^4 \over 567}t_2^2\Big[
5(1+x_2^2)\Big({655 \over 2}-78\ln 2\Big) +x_2(3187-624\ln 2)\Big]\nonumber 
\\ && + {k_f^2 \over 6} t_0 t_2(2+x_0+x_2+2x_0x_2)(4\ln 2-7)\nonumber \\ && + 
{2k_f^4 \over 63}t_1 t_2(2+x_1+x_2+2x_1x_2)(12\ln 2-31)\nonumber \\ && 
+{4k_f^4 \over 189}W_0^2 (761-132\ln 2)\bigg\}\,.\end{eqnarray} }
\noindent
Note that the third-last and second-last contribution proportional to 
$t_0t_2$ and $t_1 t_2$ are $sp$-mixing terms. These arise solely from the 
second-order Hartree diagram and they are made possible by the different 
radii of Fermi spheres involved. The reasons and conditions for 
$sp$-interference to emerge for a nuclear matter quantity are explained in 
appendix A by considering the in-medium loop for a proton-neutron pair in  
nuclear matter with unequal proton and neutron densities.

In the same way (i.e. by evaluating derivatives $\partial^2/\partial \eta^2$ at 
$\eta =0$) the second-order contribution to the spin asymmetry energy $S(k_f)$ 
of nuclear matter can be calculated. One obtains the result:
\begin{eqnarray} S(k_f)^{(\rm 2nd)} &=&{M k_f^4 \over 60\pi^4}\bigg\{ 
t_0^2\big[(1+x_0^2)(1-2\ln 2) +10x_0\big] \nonumber \\ && + {k_f^2 \over 42}
t_0 t_1\big[4(1+x_0x_1)(51-22\ln 2) +7(x_0+x_1)(57-4\ln 2)\big]\nonumber \\ && 
+{k_f^4 \over 63}t_1^2\Big[(1+x_1^2)\Big({1301 \over 6}-46\ln 2\Big) 
+x_1(607-48\ln 2)\Big]\nonumber \\ && +{k_f^4 \over 567}t_2^2\Big[
5(1+x_2^2)\Big({655 \over 2}-78\ln 2\Big) +x_2(3187-624\ln 2)\Big]\nonumber 
\\ && + {k_f^2 \over 6} t_0 t_2(2-x_0+x_2-2x_0x_2)(4\ln 2-7)\nonumber \\ && + 
{2k_f^4 \over 63}t_1 t_2(2-x_1+x_2-2x_1x_2)(12\ln 2-31)\nonumber \\ && 
+{10k_f^4 \over 189}W_0^2 (227-96\ln 2)\bigg\}\,,\end{eqnarray}
which also includes $sp$-interference terms proportional to $t_0t_2$ and 
$t_1 t_2$. When treating the spin-orbit interaction at second order, the 
spin-trace over three polarized medium-insertions eq.(7) singles out the 
third component of the vector $\vec b = \vec q_1\times \vec q_2$, but by 
spherical symmetry of the integration region one can make replacement 
$b_3^2 \to \vec b^{\,2}/3$. A good check of the intricate second-order 
calculation of asymmetry energies is provided by the relation:
\begin{equation}S(k_f)^{(\rm 2nd)}= A(k_f)^{(\rm 2nd)}_{x_{0,1}\to -x_{0,1}} - 
{Mk_f^8 \over 630\pi^4}W_0^2 (43+24\ln 2) \,,\end{equation} 
whose first part follows by working with the equivalent interaction formulated 
in terms of the isospin-exchange operator $P_\tau=(1+\vec \tau_1\!\cdot \!\vec 
\tau_2)/2$ instead of $P_\sigma=(1+\vec \sigma_1\!\cdot\!\vec\sigma_2)/2$. 
Obviously, the spin-orbit interaction produces different 
responses to asymmetries in isospin-space and spin-space. 

Finally, by reweighting the Hartree and Fock diagrams (with factors $1/2$ and 
$1$) one obtains the following second-order contribution to the spin asymmetry 
energy of pure neutron matter:
{\allowdisplaybreaks \begin{eqnarray} S_n(k_n)^{(\rm 2nd)} &=&{M k_n^4 \over 
90\pi^4}\bigg\{-t_0^2(1-x_0)^2(2+\ln 2)-{5k_n^2\over 28} t_0 t_1(1-x_0)(1-x_1)
(13+4\ln 2)\nonumber \\ && -{k_n^4\over 189}t_1^2(1-x_1)^2(130+33 \ln 2)+ 
{k_n^4\over 126} t_2^2(1+x_2)^2 (359-78 \ln 2) \nonumber \\ && + {k_n^2\over 4} 
t_0 t_2(1-x_0)(1+x_2)(4\ln 2 -7) + {k_n^4 \over 21}t_1t_2(1-x_1)(1+x_2)(12\ln 2 
- 31) \nonumber\\ &&+ {10k_n^4\over 189}W_0^2(227-96 \ln 2)\bigg\}\,.
\end{eqnarray}}
This expression includes $sp$-mixing terms and it depends on the expected 
parameter combinations. 
\section{Effective nucleon mass and Fermi-liquid parameters}
In this section, we study the effective nucleon mass and the isotropic 
Fermi-liquid parameters. Through interactions with the dense nuclear medium, 
nucleons acquire a momentum- and density-dependent self-energy and thereby
turn into quasi-particles. Long-lived quasi-particle excitations exist  
only in the vicinity the Fermi surface $|\vec p\,|=k_f$ and they are 
(kinematically) characterized by an effective mass. The ratio between the 
free nucleon mass $M$ and the density-dependent in-medium nucleon mass 
$M^*(k_f)$ can be written in the form:
\begin{equation}{M \over M^*(k_f)} = 1-{k_f^2\over 2M^2} +R(k_f)\,,
\end{equation} 
where the second term $-k_f^2/2M^2$ stems from the relativistic correction 
$-p^4/8M^3$ to the kinetic energy. The effects of interactions are represented
by the (dimensionless) quantity $R(k_f)= -2Mk_f f_1(k_f)/3\pi^2$ with 
$f_1(k_f)$ the spin- and isospin-independent $p$-wave Landau parameter 
\cite{landau}. The first order contribution from the Skyrme interaction 
$V_{\rm Sk}$ to the quantity $R(k_f)$ reads:
\begin{equation}R(k_f)^{(\rm 1st)} = {M k_f^3 \over 12\pi^2} 
\big[3t_1+t_2(5+4x_2)\big]\,. \end{equation}
In order to determine the corresponding second-order contribution, one has to 
evaluate the five direct and crossed one-loop diagrams (and their 
mirror-reflected partners) for forward scattering of two nucleons on the Fermi 
surface $|\vec p_1|=|\vec p_2|=k_f$, shown in Fig.\,2. The $p$-wave projection 
is performed by angular averaging with a weighting factor 
$3\hat p_1\!\cdot\!\hat p_2$. After a somewhat lengthy calculation one arrives 
at the following result: 
\begin{eqnarray}R(k_f)^{(\rm 2nd)} &=& {M^2 k_f^2 \over 10\pi^4} \bigg\{t_0^2(1
+x_0^2)(1-7\ln 2)+{k_f^2\over 14}t_0t_1(1+x_0 x_1)(47-104\ln 2) \nonumber \\ &&
+{k_f^4\over 378}t_1^2(1+x_1^2)(911-942\ln 2)+{k_f^4\over 1134}t_2^2(5+8x_2
+5x_2^2)(313-426\ln 2) \nonumber \\ && +{k_f^4\over 189}W_0^2(881-1032\ln 2)
\bigg\} \,, \end{eqnarray} 
which interestingly depends on the same combinations of parameters $t_0^2(1
+x_0^2)$, $t_0t_1(1+x_0 x_1)$, $t_1^2(1+x_1^2)$, $t_2^2(5+8x_2+5x_2^2)$ and 
$W_0^2$ as the second-order energy per particle $\bar E(k_f)^{(\rm 2nd)}$ written 
in eq.(16).

\begin{figure}[t]
\begin{center}
\includegraphics[scale=0.65,clip]{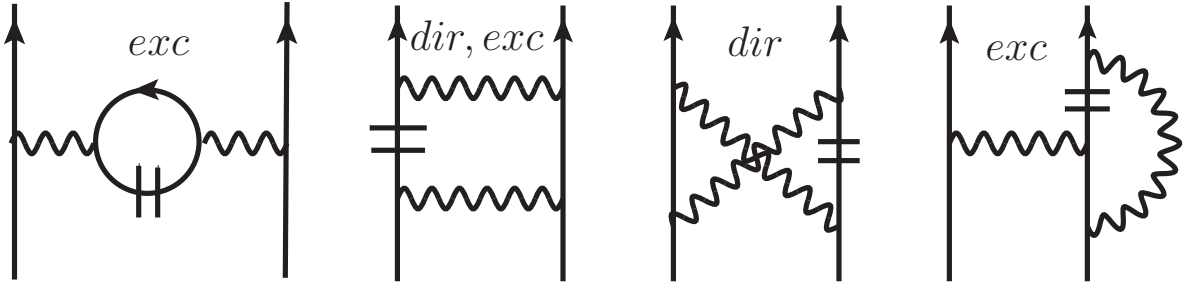}
\end{center}
\vspace{-.6cm}
\caption{Direct and exchange diagrams representing the quasi-particle 
interaction at second order. Exchange diagrams need to be 
multiplied by the negative exchange operator $-P_\tau P_\sigma$.}
\end{figure}

The isotropic (i.e. angle-averaged) quasi-particle interaction 
${\cal F}_0(k_f)$ on the Fermi surface $|\vec p_{1,2}|=k_f$ can be decomposed 
into four density-dependent ($s$-wave) Fermi-liquid parameters:
\begin{equation} {\cal F}_0(k_f) = f_0(k_f)+  f_0'(k_f)\, \vec \tau_1\!\cdot\! 
\vec \tau_2+g_0(k_f)\, \vec \sigma_1 \!\cdot\! \vec \sigma_2+  g_0'(k_f)\, 
\vec \tau_1 \!\cdot\! \vec \tau_2\, \vec \sigma_1 \!\cdot\! \vec \sigma_2\,. 
\end{equation}
The importance of the isotropic Fermi-liquid parameters $f_0(k_f)$,  
$f_0'(k_f)$ and  $g_0(k_f)$ stems from the fact that these obey exact 
relations to specific nuclear matter properties. The Landau relation 
\cite{landau} to the nuclear matter incompressibility $K(k_f)$ reads:
\begin{equation} K(k_f) = k_f^2\, {\partial^2 \bar E(k_f) \over \partial k_f^2}
+ 4k_f\, {\partial \bar E(k_f)\over \partial k_f} = {3k_f^2\over M} 
\bigg\{ 1- {k_f^2 \over 2M^2} +R(k_f) +{2 M k_f\over \pi^2} f_0(k_f) \bigg\}\,,
\end{equation}
where the expression on the right hand side is built up in the order: free 
nucleons with relativistic corrections, non-interacting quasi-nucleons, and 
finally interacting quasi-nucleons. The spin- and isospin-independent 
Fermi-liquid parameter $f_0(k_f)$ receives a first-order contribution of the 
form:  
\begin{equation}f_0(k_f)^{(\rm 1st)} = {3t_0 \over 4}+ {k_f^2 \over 8} 
\big[3t_1+t_2(5+4x_2)\big]\,, \end{equation}
and the second-order contribution derived from the one-loop scattering 
diagrams in Fig.\,2 reads: 
\begin{eqnarray}f_0(k_f)^{(\rm 2nd)} &=& {M k_f \over 4\pi^2} \bigg\{t_0^2
(1+x_0^2)(2+\ln 2)+{k_f^2\over 10}t_0 t_1(1+x_0 x_1)(41+8\ln 2) \nonumber \\ && 
+{k_f^4\over 15}t_1^2(1+x_1^2)(32+3\ln 2)+{k_f^4\over 63}t_2^2(5+8x_2+5x_2^2)
(8+3\ln 2) \nonumber \\ && +{k_f^4 \over 105}W_0^2(463+24\ln 2)\bigg\} \,. 
\end{eqnarray}
Note that in $f_0(k_f)^{(\rm 2nd)}$ the same combinations of Skyrme parameters 
appear as in eqs.(16,24). The validity of the Landau relation eq.(26) 
between $\bar E(k_f)$, $R(k_f)$ and $f_0(k_f)$ is immediately verified 
at first and second order and thus serves as an important consistency check 
of our calculation.

There exists a Landau relation \cite{landau} for the isospin asymmetry
energy $A(k_f)$ of nuclear matter:
\begin{equation} A(k_f) = {k_f^2\over 6M} 
\bigg\{ 1- {k_f^2 \over 2M^2} +R(k_f) +{2 M k_f\over \pi^2} f_0'(k_f) \bigg\}\,,
\end{equation}
which reflects again the composition: free nucleons with relativistic 
corrections, non-interacting quasi-nucleons, and interacting quasi-nucleons.  
The isospin-dependent Fermi-liquid parameter $f_0'(k_f)$ receives a 
first-order contribution of the form: 
\begin{equation}f_0'(k_f)^{(\rm 1st)} = -{t_0 \over 4}(1+2x_0)+ {k_f^2 \over 8} 
\big[t_2(1+2x_2)-t_1(1+2x_1)\big]\,, \end{equation}
and the second-order contribution derived from the one-loop scattering 
diagrams in Fig.\,2 reads: 
{\allowdisplaybreaks
\begin{eqnarray}f_0'(k_f)^{(\rm 2nd)} &=& {M k_f\over 4\pi^2} 
\bigg\{t_0^2\big[(1+x_0^2)\ln 2-2x_0\big] \nonumber \\ && +{k_f^2\over 30}t_0 
t_1\big[(1+x_0x_1)(9+32\ln 2)+(x_0+x_1)(4\ln 2-57)\big]  \nonumber \\ &&
+{k_f^4 \over 315}t_1^2\big[(1+x_1^2)(65+111\ln 2)+x_1(48\ln 2-607)\big] 
\nonumber \\ && +{k_f^4\over 63}t_2^2\big[(1+x_2^2)(19+15\ln 2)+x_2(43+24\ln 2)
\big]\nonumber \\ && +{k_f^2\over 30}t_0 t_2(2+x_0+x_2+2x_0 x_2)(4\ln 2-7)
\nonumber \\ && +{2k_f^4 \over 315}t_1 t_2(2+x_1+x_2+2x_1 x_2)(12\ln 2-31)
\nonumber \\ && +{k_f^4 \over 45}W_0^2 (103+24\ln 2) \bigg\} \,.\end{eqnarray}}
It is important to note that the Landau relation eq.(29) between 
$A(k_f)$, $R(k_f)$ and $f_0'(k_f)$ is valid at first and second order. 

In an analogous way the spin asymmetry energy $S(k_f)$ of nuclear matter is 
related to spin-spin part of the (isotropic) quasi-particle interaction:  
\begin{equation} S(k_f) = {k_f^2\over 6M} 
\bigg\{ 1- {k_f^2 \over 2M^2} +R(k_f) +{2 M k_f\over \pi^2} g_0(k_f) \bigg\}\,.
\end{equation} 
The spin-dependent Fermi-liquid parameter $g_0(k_f)$ receives from $V_{\rm Sk}$
the first-order contribution:
\begin{equation}g_0(k_f)^{(\rm 1st)} = {t_0 \over 4}(2x_0-1)+ {k_f^2 \over 8} 
\big[t_1(2x_1-1)+t_2(1+2x_2)\big]= f_0'(k_f)^{(\rm 1st)}_{x_{0,1}\to -x_{0,1}}\,, 
\end{equation}
and the second-order contribution derived from the one-loop scattering 
diagrams in Fig.\,2 reads:
\begin{equation}g_0(k_f)^{(\rm 2nd)} = f_0'(k_f)^{(\rm 2nd)}_{x_{0,1}\to -x_{0,1}}- 
{M k_f^5 \over 210 \pi^2} W_0^2(43+24\ln 2) \,, \end{equation}
with $f_0'(k_f)^{(\rm 2nd)}$ written in eq.(31). The Landau relation eq.(32) 
between $S(k_f)$, $R(k_f)$ and $g_0(k_f)$ is just as well satisfied at first 
and second order. 

Finally, there is the isovector spin-spin Fermi-liquid parameter $g_0'(k_f)$. 
The corresponding contributions from the Skyrme interaction $V_{\rm Sk}$ at 
first and second order read:
\begin{equation}g_0'(k_f)^{(\rm 1st)} = -{t_0 \over 4}+ {k_f^2 \over 8}(t_2-t_1)
\,, \end{equation}
\begin{eqnarray}g_0'(k_f)^{(\rm 2nd)} &=& {M k_f \over 12\pi^2} \bigg\{t_0^2\big[
3\ln 2-x_0^2(2+\ln 2)\big] +{k_f^2\over 10}t_0 t_1\big[
9+32\ln 2 -x_0 x_1(41+8\ln 2)\big]  \nonumber \\ &&+{k_f^4 \over 105}
t_1^2\big[65+111\ln 2-7x_1^2(32+3\ln 2)\big] \nonumber \\ && +{k_f^4 \over 21}
t_2^2\big[19+22x_2+8x_2^2+3(5+8x_2+x_2^2)\ln 2\big]+{k_f^2\over 5}t_0 t_2
(1+x_2)(4\ln 2-7)\nonumber \\ && +{4k_f^4 \over 105}t_1 t_2(1+x_2)(12\ln 2-31) 
+{k_f^4 \over 45}W_0^2 (103+24\ln 2) \bigg\} \,. \end{eqnarray}
Note that $sp$-mixing terms are present in the three spin- or isospin-dependent 
Fermi-liquid parameters. 

Additional checks of our calculation are provided 
by the Fermi-liquid parameters $f_{n0}(k_n)$ and  $g_{n0}(k_n)$ of neutrons in 
pure neutron matter. In order to obtain these from the computation of the 
isotropic quasi-particle interaction ${\cal F}_0(k_f)$ in isospin-symmetric 
nuclear matter, one merely has to set $P_\tau = 1$, replace $k_f \to k_n$, and 
multiply the left diagram in Fig.\,2 (with a closed nucleon ring) by a factor 
$1/2$. The $p$-wave Landau parameter $f_{n1}(k_n)$ enters the effective neutron 
mass ratio $R_n(k_n)$ with a factor $-M k_n/3\pi^2$. For the sake of 
completeness we list the relevant expressions for the neutron 
quasi-particle properties at first order:
\begin{equation} R_n(k_n)^{(\rm 1st)} = {M k_n^3 \over 12\pi^2}\big[ t_1(1-x_1)+3 
t_2(1+x_2)\big]\,,\end{equation}
 \begin{equation} f_{n0}(k_n)^{(\rm 1st)} = {t_0\over 2}(1-x_0) +{k_n^2 \over 4}
\big[ t_1(1-x_1)+3 t_2(1+x_2)\big]\,, \end{equation}
 \begin{equation} g_{n0}(k_n)^{(\rm 1st)} = {t_0\over 2}(x_0-1) +{k_n^2 \over 4}
\big[ t_1(x_1-1)+t_2(1+x_2)\big]\,, \end{equation}
and at second order:
\begin{eqnarray}R_n(k_n)^{(\rm 2nd)} &=& {M^2 k_n^2 \over 30\pi^4} \bigg\{t_0^2(1
-x_0)^2(1-7\ln 2)+{k_n^2\over 14}t_0t_1(1-x_0)(1-x_1)(47-104\ln 2) \nonumber \\ 
&&+{k_n^4\over 378}t_1^2(1-x_1)^2(911-942\ln 2)+{k_n^4\over 126}t_2^2(1+x_2)^2
(313-426\ln 2) \nonumber \\ && +{2k_n^4\over 189}W_0^2(881-1032\ln 2)
\bigg\} \,, \end{eqnarray} 
\begin{eqnarray}f_{n0}(k_n)^{(\rm 2nd)} &=& {M k_n \over 6\pi^2} \bigg\{t_0^2
(1-x_0)^2(2+\ln 2)+{k_n^2\over 10}t_0t_1(1-x_0)(1-x_1)(41+8\ln 2)\nonumber \\ && 
+{k_n^4\over 15}t_1^2(1-x_1)^2(32+3\ln 2)+{k_n^4\over 7}t_2^2(1+x_2)^2
(8+3\ln 2)  \nonumber \\ && +{2k_n^4 \over 105}W_0^2(463+24\ln 2)\bigg\} \,. 
\end{eqnarray}
\begin{eqnarray}g_{n0}(k_n)^{(\rm 2nd)} &=& {M k_n\over 6\pi^2} 
\bigg\{t_0^2(1-x_0)^2(\ln 2-1) +{2k_n^2\over 5}t_0 t_1(1-x_0)(1-x_1)(3\ln 2-4)
\nonumber \\ &&+{k_n^4 \over 70}t_1^2(1-x_1)^2(30\ln 2-53)+{3k_n^4\over 14}
t_2^2(1+x_2)^2(3+2\ln 2)\nonumber \\ && +{k_n^2\over 10}t_0 t_2(1-x_0)(1+x_2)
(4\ln 2-7)+{2k_n^4 \over 105}t_1 t_2(1-x_1)(1+x_2)(12\ln 2-31)\nonumber \\ && 
+{2k_n^4 \over 315}W_0^2 (463+24\ln 2) \bigg\} \,.\end{eqnarray}
The pertinent Landau relations to the incompressibility $K_n(k_n)$ and spin 
asymmetry energy $S_n(k_n)$ of pure neutron matter (analogous to eqs.(26,32) 
but with a prefactor $Mk_n/\pi^2$ at the last summand) are satisfied at 
first and second order. The quasi-particle interaction of neutrons in pure 
neutron matter, taking into account also the non-central tensor and 
cross-vector terms, has been calculated in ref.\cite{holt} on the basis of 
chiral two- and three-nucleon potentials in many-body 
perturbation theory up to second order.
\section{Tensor interactions}
The Skyrme interaction $V_{\rm Sk}$ in eq.(1) depends on seven parameters. This
are two less than the nine low-energy constants $C_{S,T}$ and $C_j\,, j=1,...,7$ 
appearing in general NN contact-potential of chiral effective field theory 
\cite{evgeni,hammer,machleidt} up to order $p^2$. A complete matching a 
achieved by supplementing to the Skyrme interaction $V_{\rm Sk}$ two additional 
tensor terms of the Galilei- and time-reversal-invariant form:
 \begin{eqnarray} V_{\rm ten} &=& t_4\big[ 3\,\vec \sigma_1\!\cdot\!\vec q_{\rm out}
\,\vec \sigma_2\!\cdot\!\vec q_{\rm out}+3\,\vec \sigma_1\!\cdot\! \vec q_{\rm in}\,
\vec \sigma_2\!\cdot\!\vec q_{\rm in} - \vec \sigma_1\!\cdot\!\vec \sigma_2
(\vec q_{\rm out}^{\,2}+ \vec q_{\rm in}^{\,2})\big] \nonumber \\ && + t_5\big[ 
3\,\vec \sigma_1\!\cdot\! \vec q_{\rm out}\,\vec \sigma_2\!\cdot\! \vec q_{\rm in} 
+3\,\vec \sigma_1\!\cdot\! \vec q_{\rm in}\,\vec \sigma_2\!\cdot\! \vec 
q_{\rm out} - 2\,\vec \sigma_1\!\cdot\!\vec \sigma_2 \,\vec q_{\rm out}\!\cdot 
\vec q_{\rm in}\big]\,.\end{eqnarray}
Note that for both terms in $V_{\rm ten}$ the spin-matrices $\sigma_1^i\sigma_2^j$ 
are contracted with symmetric and traceless tensors depending on the external 
momentum-differences $\vec q_{\rm in}$ and $\vec q_{\rm out}$. Because of these 
properties the first-order contribution from $V_{\rm ten}$ as well as all its
interference terms with $V_{\rm Sk}$ vanish, regarding energies per particle 
and the central quasi-particle interaction in nuclear matter.

Using dimensional regularization (as discussed in the beginning of section\,3) 
and evaluating the second-order Hartree and Fock diagrams in Fig.\,1 with the 
tensorial contact-interaction $V_{\rm ten}$, one obtains the following 
contributions to the energy per particle of isospin-symmetric nuclear matter:
 \begin{equation}\bar E(k_f)^{\rm(ten)} = {M k_f^8 \over 15 \pi^4} \bigg\{ 
{t_4^2 \over 231} (1525-129\ln 2) +t_5^2 (19-3 \ln 2)\bigg\}\,,\end{equation}
and to the energy per particle of pure neutron matter:
\begin{equation}\bar E_n(k_n)^{\rm(ten)} = {2M k_n^8 \over 45 \pi^4}\, t_5^2 
(19-3 \ln 2)\,.\end{equation}
 
The asymmetric medium-insertions introduced in eqs.(4,7) lead to the 
following second-order contributions from $V_{\rm ten}$ to the isospin 
asymmetry energy of nuclear matter: 
\begin{equation} A(k_f)^{\rm(ten)} = {M k_f^8 \over 945 \pi^4}\bigg\{ t_5^2 
(3719-708 \ln 2) -t_4^2\Big( 100\ln 2+{317 \over 3}
\Big)+24\,t_4 t_5(12\ln 2 -31)  \bigg\} \,,\end{equation}
and to the spin asymmetry energy of nuclear matter:
\begin{equation} S(k_f)^{\rm(ten)} = {M k_f^8 \over 54 \pi^4}\bigg\{ {11t_4^2 
\over 15}(13-48\ln 2)+{t_5^2\over 7}(395-1104\ln 2)\bigg\}\,.\end{equation}
Interestingly, the $t_4t_5$-interference term is absent in $\bar E(k_f)^{\rm(ten)}$
and $S(k_f)^{\rm(ten)}$. The actual calculation of the second-order Hartree 
diagram produces in both cases an integrand $H=t_4t_5(|\vec q_1|^4-|\vec q_2|^4
)$ for the triple Fermi-sphere integral in eq.(15), but due to the symmetry 
$\vec q_1 \leftrightarrow \vec q_2$ of the other ingredients it integrates to 
zero. By reweighting the Hartree diagram with a factor $1/2$, the 
second-order contribution from $V_{\rm ten}$ to the spin asymmetry 
energy of pure neutron matter follows as:  
\begin{equation} S_n(k_n)^{\rm(ten)} = {M k_n^8 \over 567 \pi^4}\,t_5^2
(395-1104\ln 2)\,.\end{equation}
One observes that the neutron matter quantities written in eqs.(39,42) depend 
only on the parameter $t_5$. This feature is explained by the fact the first 
tensor term in eq.(37) proportional to $t_4$ is responsible for the mixing 
between the $^3\!S_1$-state and $^3\!D_1$-state with total isospin $I=0$. 

We continue with presenting the second-order contributions from $V_{\rm ten}$   
to the effective nucleon mass ratio:
\begin{equation} R(k_f)^{\rm(ten)} ={M^2 k_f^6\over 126\pi^4} \bigg\{{t_4^2\over 
3} (929-888 \ln 2)+{t_5^2\over 5}(5147-6504 \ln 2)\bigg\}\,,\end{equation}
and to the isotropic Fermi-liquid parameters:
\begin{equation} f_0(k_f)^{\rm(ten)} ={M k_f^5\over 140\pi^2} \bigg\{t_4^2 \Big(
{2195 \over 3} + 88 \ln2\Big) + t_5^2 (2029 + 312 \ln2)\bigg\}\,,\end{equation}
\begin{equation} f'_0(k_f)^{\rm(ten)} ={M k_f^5\over 140\pi^2} \bigg\{3t_4^2(
40\ln2-73) + {32\over 3} t_4t_5(12\ln2 -31)+t_5^2(1081+408\ln2)\bigg\}\,,
\end{equation}
\begin{equation} g_0(k_f)^{\rm(ten)} =-{M k_f^5\over 20\pi^2} \bigg\{{t_4^2\over 
63}(881+984 \ln2)+ t_5^2(19 + 72 \ln2)\bigg\}\,,\end{equation}
\begin{equation} g'_0(k_f)^{\rm(ten)} ={M k_f^5\over 140\pi^2} \bigg\{t_4^2
(40\ln2-73) + {32\over 9} t_4t_5(31-12\ln2)-{t_5^2\over 9}(1613+2424\ln2)
\bigg\}\,.\end{equation}
Note that $t_4t_5$-interference terms are present only in the isovectorial 
quantities $f'_0(k_f)$ and  $g'_0(k_f)$. The simultaneous validity of the three 
Landau relations in eqs.(26,29,32) provides an excellent check of our 
second-order calculation with the zero-range tensor interaction $V_{\rm ten}$.   
The second-order contributions from $V_{\rm ten}$ to the neutron quasi-particle 
properties are obviously proportional to $t_5^2$ and they read:   
\begin{equation} R_n(k_n)^{\rm(ten)} = {M^2 k_n^6\over 945\pi^4}\,t_5^2(5147-6504 
\ln 2)\,, \end{equation}
\begin{equation}  f_{n0}(k_n)^{\rm(ten)} ={M k_n^5\over 105\pi^2} \,t_5^2 
(2029 + 312 \ln2)\,, \qquad g_{n0}(k_n)^{\rm(ten)} =-{M k_n^5\over 15\pi^2}\, t_5^2 
(19 + 72 \ln2)\,. \end{equation}

\section{Iteration of {\boldmath$V_{\rm Sk}$} with one-photon 
exchange}
\begin{figure}
\begin{center}
\includegraphics[scale=0.65,clip]{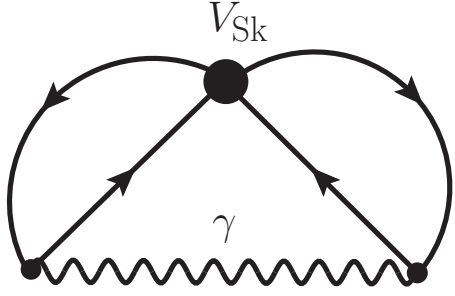}
\end{center}
\vspace{-.6cm}
\caption{Iteration of the zero-range Skyrme interaction with one-photon 
exchange.}
\end{figure}

The Skyrme interaction $V_{\rm Sk}$ in eq.(1) can be viewed as a suitable
phenomenological parameterization of the strong nuclear force. In nuclear 
matter (or finite nuclei) it affects protons and neutrons in the same way. 
This isospin-degeneracy gets lifted by taking into account the Coulomb 
interaction between the protons. A further charge-symmetry breaking 
contribution is provided by iterating the (strong) Skyrme interaction 
$V_{\rm Sk}$ with the (electromagnetic) one-photon exchange, as depicted by the 
three-loop diagram in Fig.\,3. In the case at hand the iterated diagram with 
two medium-insertions does not vanish in dimensional regularization due to the 
presence of a photon-propagator $[m_\gamma^2+(\vec l \pm \vec q\,)^2]^{-1}$, with 
$m_\gamma$ an infinitesimal photon mass. The limit $m_\gamma \to 0 $ has to be 
taken after performing the $d^3l$-loop integral, which results in: 
$\arctan(2|\vec q\,|/m_\gamma) \to \pi/2$.  After combination with the 
Pauli-blocking corrections (supplied by the iterated diagram with three 
medium-insertions) one obtains the following contribution to the energy per 
proton:    
 \begin{eqnarray} \bar E_p(k_p) &=& {\alpha M k_p^2 \over 10 \pi^3} 
\bigg\{ t_0(1-x_0)\Big(1-{\pi^2 \over 3} -2 \ln 2\Big) + {3k_p^2 \over 14}  
 \bigg[t_1(1-x_1)\Big({31 \over 4}-{\pi^2\over 3} -3 \ln 2\Big)\nonumber 
\\ && +t_2(1+x_2)\Big({5 \over 4}-\pi^2 -5 \ln2\Big)+{2\mu_p^2
\over 3M^2}\,t_0(1-x_0)(11-2\ln2)\bigg]\bigg\}\,,
\end{eqnarray}
where $\alpha = e^2/4\pi= 1/137.036$ denotes the electromagnetic fine-structure 
constant and the proton Fermi momentum $k_p$ is related to the proton density 
by $\rho_p=k_p^3/3\pi^2$. Remarkably, the result for $\bar E_p(k_p)$ is 
infrared-finite, since an infrared regulator term $\ln(k_p/m_\gamma)$ drops out 
in the end by angular integration: $\int_{-1}^1\!dy\,[f(y)\!-\!f(-y)]=0$. The 
last term in eq.(56) comes from iterating the $s$-wave contact part $t_0
(1-x_0)$ with the magnetic interaction $\sim \vec \sigma\!\times\!\vec q_j$, 
where $\mu_p = 2.79$ denotes the proton magnetic moment in units of the 
nuclear magneton $e/2M$. For the sake of completeness, we give also the 
contribution from the iterated diagram in Fig.\,3 to the spin asymmetry energy 
of protons:    
\begin{eqnarray} S_p(k_p) &=& {\alpha M k_p^2 \over 18 \pi^3} 
\bigg\{ t_0(1-x_0)\Big({2\pi^2 \over 3} +1 -2 \ln 2\Big) + {k_p^2 \over 5}  
 \bigg[t_1(1-x_1)\Big({2\pi^2\over 3} -{47 \over 4} + \ln 2\Big)\nonumber \\ 
&& -t_2(1+x_2)\Big({4\pi^2\over 3}+{5 \over 4} +5 \ln2\Big)+{4\mu_p^2
\over M^2}\,t_0(x_0-1)(2+\ln2)\bigg]\bigg\}\,.\end{eqnarray}
As an aside we note that the one-photon exchange with the magnetic coupling 
gives also (small) first-order Fock contributions of the form: 
$\bar E_p(k_p)=-3 S_p(k_p)=\alpha\,\mu_p^2 k_p^3/(6\pi M^2)$.  

The charge-symmetry breaking contribution $\rho_p \bar E_p(k_p)$ to the energy 
density of protons in nuclei could play a role in resolving the 
Nolen-Schiffer anomaly \cite{nolen,shlomo}. This anomaly refers to the 
observation that the binding energy differences of mirror nuclei cannot be 
explained by the Coulomb interaction of protons alone, using realistic nuclear 
wave functions. In a systematic study by Brown et al.\,\cite{brown} it was 
found that the anomaly can be resolved by either dropping the attractive 
Coulomb Fock-term $\bar E_p(k_p)= -3\alpha k_p/4\pi$ or by introducing a 
charge-symmetry breaking delta-force. The first term in eq.(56) proportional 
to $k_p^2$ has been investigated in ref.\,\cite{pigamma}  with respect to its 
capability of compensating the attractive Coulomb Fock-term. Such a study 
should be extended to the iteration of the complete Skyrme interaction 
$V_{\rm Sk}$ with the one-photon exchange. Actually, the novel charge-symmetry 
breaking contribution $\bar E_p(k_p)$ written in eq.(56) should be included 
routinely in Skyrme calculations of medium-mass and heavy nuclei.

\section{Provisional parameter fits}
\begin{figure}[h!]
\begin{center}
\includegraphics[scale=0.5,clip]{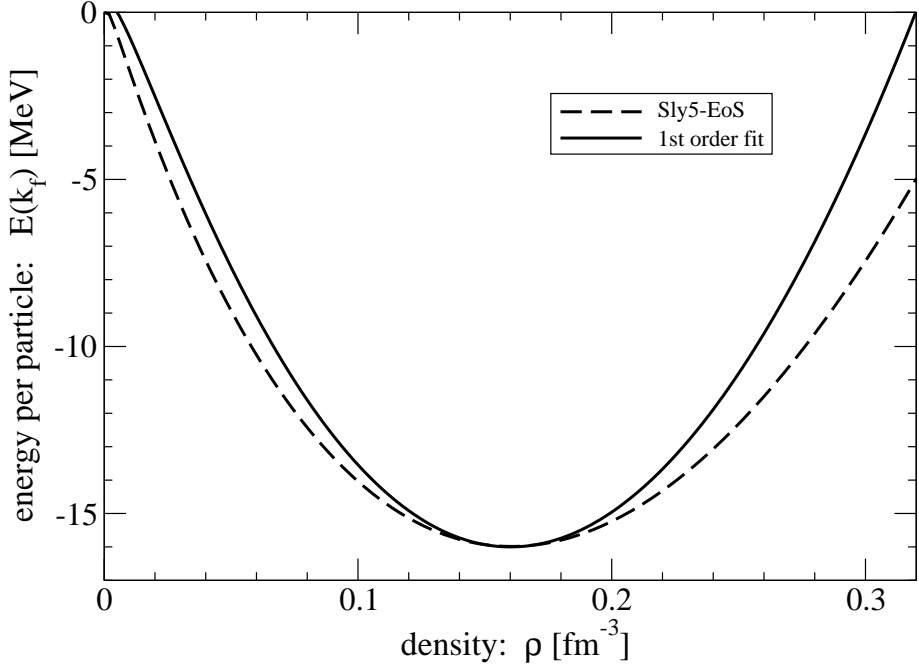}
\end{center}
\vspace{-.6cm}
\caption{Saturation curve of isospin-symmetric nuclear matter. The dashed line 
stems from the SLy5-interaction \cite{sly5}. The full curve is a 
first-order fit without the density-dependent term ${1 \over 6}t_3\rho^{1/6}$.}
\end{figure}

In this section we perform some provisional parameter fits in order to learn 
about possible improvements through the inclusion of second-order contributions 
from $V_{\rm Sk}$. We restrict the discussion to the nuclear matter saturation 
curve in the density region $0<\rho<2\rho_0=0.32\,$fm$^{-3}$. The equation of 
state provided by the phenomenological SLy5-interaction \cite{sly5} is 
taken as a benchmark. Out of interest we perform first a fit to the energy per 
particle at first order: $\bar E(k_f)= \bar E(k_f)^{(\rm kin)}-a_3 k_f^3+a_5k_f^5$. 
The empirical nuclear matter saturation point ($k_{f0}=263\,$MeV, $\bar E_0=
-16\,$MeV) fixes the two free parameters to the values: $a_3=30.81\,$MeVfm$^3$, 
 $a_5 = 8.356\,$MeVfm$^5$, and the resulting saturation curve is shown by the 
full line in Fig.\,4. It is considerably stiffer than the realistic equation of 
state (represented by the dashed line in Fig.\,4) as evidenced by the high 
value $K=307\,$MeV of the nuclear matter incompressibility. While the shape 
of the full line in Fig.\,4 can be considered as acceptable, there is serious 
problem behind it, namely the unrealistically low value of effective nucleon 
mass $M^*(k_{f0})= 0.39M$, which is completely determined by the parameter 
$a_5$. This (unwanted) rigid correlation at first order is usually avoided by 
introducing the density-dependent term ${1\over 6}t_3(1+x_3 P_\sigma)\rho^{1/6}$.
     
Next, we include in the fit the complete second-order contribution $\bar 
E(k_f)^{\rm (2nd)}+\bar E(k_f)^{\rm (ten)}$  written in eqs.(16,44). The nine 
parameters are determined 
by minimizing the squared deviations from the realistic nuclear matter 
saturation curve at about 40 equally-spaced points in the density region 
$0<\rho<2\rho_0=0.32\,$fm$^{-3}$. The resulting fit values are: $t_0 = 
-3.535\,$fm$^2$, $t_1 = 1.156\,$fm$^4$, $t_2 = -3.942\,$fm$^4$, and the others 
very close to zero: $x_{0,1,2}=0$, $t_{4,5}=0$, $W_0=0$. An inspection of the 
full line in Fig.\,5 reveals that a good reproduction of the realistic nuclear 
matter saturation curve cannot be obtained. The reason for this failure lies 
presumably in the too strong density-dependence $k_f^8 \sim \rho^{8/3}$ of 
several second-order contributions. But interestingly with the above parameter 
values, the isospin asymmetry energy $A(k_{f0}) =S(k_{f0}) = 33.0\,$MeV comes 
out close to its empirical value. The previous free fit gave zero even for the 
spin-orbit parameter $W_0$. If one fixes it to its (large) empirical value 
$W_0 = 126\,{\rm MeVfm}^5= 0.638\,$fm$^4$ \cite{sly5}, the other parameters are 
fitted to: $t_0 = -3.849\,$fm$^2$, $t_1=1.778\,$fm$^4$,  $t_2=-3.341\,$fm$^4$,
and $x_{0,1,2}=0$, $t_{4,5}=0$. The resulting equation of state is shown by the 
dashed-dotted line in Fig.\,5 and one observes that its shape remains almost 
unchanged in comparison to the free fit. The values for the isospin asymmetry 
energy $A(k_{f0}) = 36.0\,$MeV and the spin asymmetry energy $S(k_{f0}) = 
32.3\,$MeV have changed a bit.

From these findings one can conclude that the inclusion of the (ad hoc)
density-dependent term  ${1\over 6}t_3(1+x_3 P_\sigma) \rho^{1/6}$ is 
indispensable for a realistic description of nuclear matter in the Skyrme 
approach. Formally, this extra term is taken into account by making the 
parameter substitutions: $t_0\to t_0+{1\over 6}t_3\rho^{1/6}$ and  $t_0x_0\to 
t_0x_0 +{1\over 6}t_3x_3 \rho^{1/6} $, in all first- and second-order 
expressions. We have convinced ourselves that after this extension a good 
reproduction of the realistic nuclear matter saturation curve is possible at 
second order. This is in agreement with the findings of refs.\,\cite{moghrabi2,
moghrabi3}, where however incorrect expressions have been used for the 
second-order contributions. Certainly, the fitting of parameters of the Skyrme 
interaction $V_{\rm Sk }$, when treated to second order, should go beyond the 
mere saturation curve and include as constraints all possible available 
properties of nuclear matter. Such a comprehensive analysis goes beyond the 
scope of the present paper. But the analytical calculations of a large variety 
of nuclear matter properties to first and second order as performed in this 
work lay the foundation for such a future study.     
\begin{figure}[h!]
\begin{center}
\includegraphics[scale=0.5,clip]{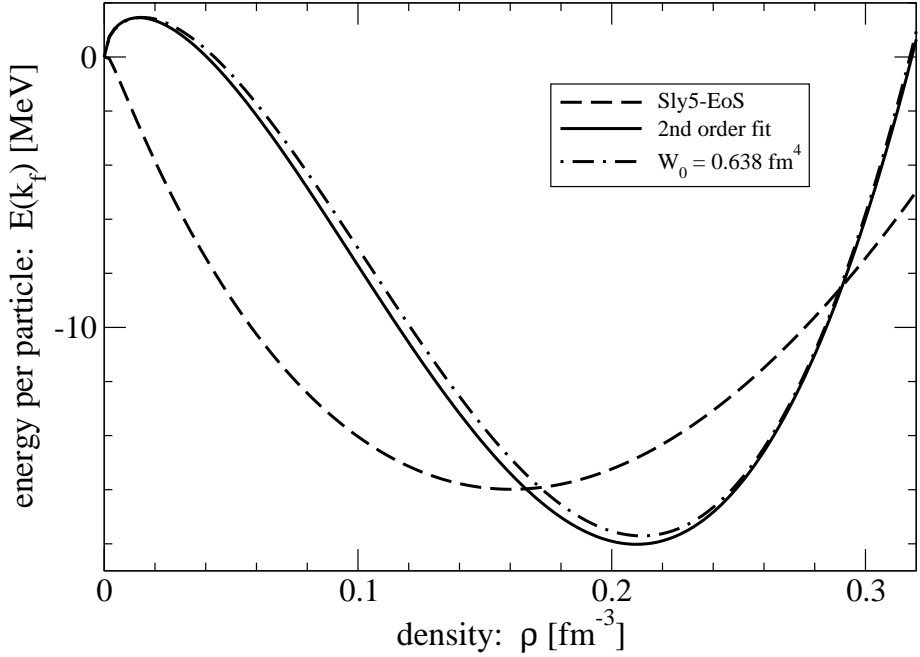}
\end{center}
\vspace{-.6cm}
\caption{Fits to the nuclear matter saturation curve with inclusion of 
second-order contributions from the Skyrme interaction $V_{\rm Sk}$. The 
(ad hoc) density-dependent term ${1 \over 6}t_3\rho^{1/6}$ is left out.}
\end{figure}

\section*{Appendix A: In-medium loop featuring s-wave and 
p-wave mixing}
\begin{figure}[h!]
\begin{center}
\includegraphics[scale=0.7,clip]{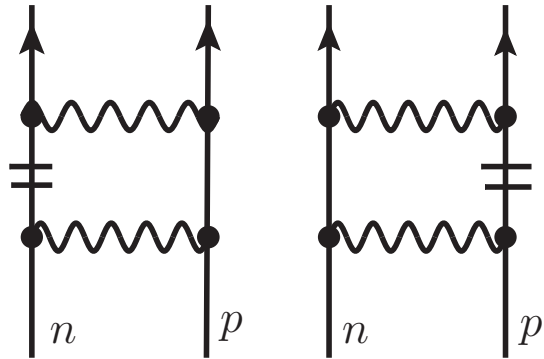}
\end{center}
\vspace{-.6cm}
\caption{In-medium loop for an interacting proton-neutron pair. The short 
double-lines symbolize Pauli-blocking effects in the nuclear medium.}
\end{figure}
In this appendix we analyze under which conditions a mixing between $s$-wave 
and $p$-wave components of the two-body interaction (with different parity)
can occur in a fermionic medium. The two diagrams in Fig.\,6 represent the 
in-medium loop for an interacting proton-neutron pair in nuclear matter. The 
short double-line symbolizes the medium-insertion (i.e. the Pauli-blocking 
effect) for the neutron on the left side and for the proton on the right 
side. If the lower and upper contact-vertex are of $s$-wave type and $p$-wave 
type, respectively, the real part of the in-medium loop is given by a 
principal-value integral of the form: 
\begin{eqnarray} && -\hspace{-0.45cm}\int\!{d^3 l \over (2\pi)^3}\,{
(\vec l^{\,2})^j\,\vec l\!\cdot\! \vec q \over \vec l^{\,2}-\vec q^{\,2}}\Big\{
\theta(k_n-|\vec P-\vec l\,|)+\theta(k_p-|\vec P+\vec l\,|)\Big\} \nonumber 
\\ && =  -\hspace{-0.45cm}\int\!{d^3 l \over 
(2\pi)^3}\,{(\vec l^{\,2})^j\,\vec l\!\cdot \!\vec q \over \vec l^{\,2}
-\vec q^{\,2}}\Big\{\theta(k_n-|\vec P-\vec l\,|)-\theta(k_p-|\vec P-
\vec l\,|)\Big\} \nonumber \\ && =\vec P \!\cdot \!\vec q\,\, {k_f^{2j+1}\delta 
\over 24\pi^2}\, I_j(s,\kappa)+{\cal O}(\delta^3)\,,\end{eqnarray}
where $k_{n,p}= k_f(1\pm \delta)^{1/3}$ denote the Fermi momenta of neutrons and 
protons, and $j=0,1$ are exponents of $\vec l^{\,2}$. The expression in the
second line follows by substituting $\vec l\to-\vec l$ in the second 
summand. It makes evident that $sp$-mixing can occur only in a medium with 
unequal neutron and proton densities, $k_n \ne k_p$. For small 
isospin asymmetries $\delta$ the available phase-space becomes a (thin) 
spherical shell of thickness $2k_f \delta/3$. The net result for the 
in-medium loop with $sp$-mixing, expanded to linear order in $\delta$, is given 
in the third line of eq.(58) and it involves the logarithmic functions:
\begin{equation}I_0(s,\kappa) = {1\over s^3} \bigg\{ 4s+(s^2+\kappa^2-1) \ln
{|(s+1)^2-\kappa^2| \over |(s-1)^2-\kappa^2|} \bigg\} \,,\qquad
I_1(s,\kappa) = 8 +\kappa^2 I_0(s,\kappa)\,,\end{equation}
where $s=|\vec P|/k_f$ and $\kappa =|\vec q\,|/k_f$ are two dimensionless 
variables. In fact, the same asymmetric configuration as considered above is 
present in the calculation of the isospin (or spin) asymmetry energy from the 
second-order Hartree and Fock diagrams. This explains why $sp$-interference 
terms proportional to $t_0t_2$ and $t_1t_2$ do occur in the expressions 
$A(k_f)^{\rm (2nd)}$, $S(k_f)^{\rm(2nd)}$ and $S_n(k_n)^{\rm (2nd)}$  written in 
eqs.(18,19,21). In the case of the energy per particle, $\bar E(k_f)$ or  
$\bar E_n(k_n)$, one has from the outset just one single Fermi momentum, 
$k_f$ or $k_n$, and therefore $sp$-mixing terms cannot emerge for these 
quantities at second order in many-body perturbation theory.

\section*{Appendix B: Interaction through long-range potentials}
In this appendix we derive the first-order Fock and second-order Hartree 
and Fock contributions as they arise from several long-range interaction 
potentials. The energy per particle and asymmetry energies are calculated for 
many-fermion systems with either four or two internal degrees of freedom 
(e.g. isospin-symmetric nuclear matter or pure neutron matter). Our first 
candidate is an interaction potential in coordinate-space with an inverse 
square and an inverse fourth-power dependence on the particle-distance:
\begin{equation} \widetilde V_{\rm long} = {\beta_2 \over |\vec r_1\!-\!
\vec r_2|^2}+{\beta_4 \over |\vec r_1\!-\!\vec r_2|^4}\,,\end{equation}
where the parameter $\beta_2$ is a length and $\beta_4$ is a volume. The 
calculation of interaction contributions in the fermionic medium requires  
the Fourier-transform of this long-range potential:
\begin{equation} V_{\rm long} = {2\pi^2 \beta_2 \over |\vec q_{\rm out}\!-\!\vec 
q_{\rm in}|}-\pi^2 \beta_4 |\vec q_{\rm out}\!-\!\vec q_{\rm in}|\,,\end{equation}
where $\vec q_{\rm out}\!-\!\vec q_{\rm in}$ is the Galilei-invariant momentum 
transfer. Taking $V_{\rm long}$ as an interaction-vertex in momentum-space, the 
evaluation of the first-order Fock diagram and the second-order Hartree and Fock 
diagrams in Fig.\,1 gives for the four-component system (nuclear matter):
\begin{eqnarray} \bar E(k_f) &=& -{\beta_2 k_f^2 \over 5}  +{3\beta_4 k_f^4
\over 35} +  {\beta_2^2 M k_f^2 \over 5}\Big( 2 G -{\pi^2 \over 3}-2\ln 2\Big) 
\nonumber \\ && + {\beta_2\beta_4 M k_f^4 \over 35}(4\ln 2-13 -6 G) + {\beta_4^2 
M k_f^6\over 315}\Big(5G+{731\over 18}-8\ln 2\Big)\,,\end{eqnarray}
\begin{eqnarray}  A(k_f)=S(k_f) &=& -{\beta_2 k_f^2 \over 9}+{2\beta_4 k_f^4
\over 15} +  {\beta_2^2 M k_f^2 \over 9}\Big( 2 G +{\pi^2 \over 6}-2\ln 2\Big) 
\nonumber \\ && + {4\beta_2\beta_4 M k_f^4 \over 45}(1-3 G+2\ln 2) + {\beta_4^2 
M k_f^6\over 105}\Big(5G+{103\over 9}-8\ln 2\Big)\,,\end{eqnarray}
where $G = \int_0^1\!dx\,x^{-1}\arctan x = \sum_{n=0}^\infty (-1)^n(2n+1)^{-2} = 
0.9159656$ denotes the Catalan number. Note that $G$ always results from the 
$d^3l$-loop integral belonging to  the second-order Fock diagram with two 
medium-insertions. The analogous calculation in the two-component system 
(pure neutron matter) gives the following contributions to the energy per 
particle and the spin asymmetry energy:
\begin{eqnarray}  \bar E_n(k_n) &=& -{\beta_2 k_n^2 \over 5}  +{3\beta_4 k_n^4
\over 35} + {\beta_2^2 M k_n^2 \over 10}\Big( 4 G -{\pi^2 \over 3}-1-2\ln 2\Big) 
\nonumber \\ && + {2\beta_2\beta_4 M k_n^4 \over 35}(\ln 2-1-3 G) +{\beta_4^2 M 
k_n^6\over 315}\Big(5G+{115\over 9}-4\ln 2\Big)\,,\end{eqnarray}
\begin{eqnarray}  S_n(k_n) &=& -{\beta_2 k_n^2 \over 9}+{2\beta_4 k_n^4\over 15} 
+ {\beta_2^2 M k_n^2 \over 18}\Big( 4 G +{\pi^2 \over 6}-1-2\ln 2\Big) 
\nonumber \\ && + {\beta_2\beta_4 M k_n^4 \over 45}(11-12 G+4\ln 2) + {\beta_4^2 
M k_n^6\over 105}\Big(5G-{65\over 36}-4\ln 2\Big)\,.\end{eqnarray}
It is worth to mention that the (singular) $\beta_2$-interaction provides an 
infrared-finite result at second order. Ultraviolet divergences have been 
treated in dimensional regularization, using the rule: $\int_0^\infty \! dl\,l^n 
= 0$. Note that the contributions of second-order Hartree diagrams (with two 
or three medium insertions) are in nuclear matter a factor $2$ larger than in 
pure neutron matter. The corresponding pieces carry (besides a rational number) 
the coefficients $\pi^2/3$ and $\ln 2$.
     
As a second example for a long-range interaction we treat the familiar 
Van-der-Waals potential:
\begin{equation} \widetilde V_{\rm vdw} = {12B \over |\vec r_1\!-\!
\vec r_2|^6}\,, \qquad V_{\rm vdw} = \pi^2 B |\vec q_{\rm out}\!-\!
\vec q_{\rm in}|^3\,,\end{equation}
where the factor $12$ has been introduced in order to get a simpler 
Fourier-transform. The many-body calculation at first and second order with the 
cubic interaction-vertex $ \pi^2B|\vec q_{\rm out}\!-\!\vec q_{\rm in}|^3$  leads 
to the following results for the four-component system:
\begin{eqnarray} \bar E(k_f) &=& -{8B k_f^6 \over 63}+{B^2M k_f^{10}\over 2145}
\bigg({1115603 \over 1050} +27 G -128 \ln 2\bigg)\nonumber \\ && +{2B\beta_2 
M k_f^6\over 315}(30G+53-16\ln2)+{B\beta_4 Mk_f^8 \over 231}\Big(16\ln2-7G- 
{3133\over 30}\Big) \,, \end{eqnarray}
\begin{eqnarray} A(k_f) = S(k_f) &=&-{8B k_f^6 \over 21} +{B^2 M k_f^{10} 
\over297}\bigg({613343 \over 1050} +27 G -128 \ln 2\bigg)\nonumber \\ && 
+{4B\beta_2 M k_f^6\over 315}(45G-8-24\ln2)+{4B\beta_4 Mk_f^8 \over 189}\Big(
16\ln2-7G- {1369\over 30}\Big) \,, \end{eqnarray}
and for the two-component system: 
\begin{eqnarray} \bar E_n(k_n) &=& -{8B k_n^6 \over 63}+{B^2 M k_n^{10}\over 195}
\bigg({14311 \over 350}+{27G-64\ln 2\over 11}\bigg)\nonumber \\ && +{4B\beta_2 
M k_n^6\over 945}(45G-4-12\ln2)+{B\beta_4 Mk_n^8 \over 231}\Big(8\ln2-7G- 
{1141\over 30}\Big) \,,\end{eqnarray}
\begin{eqnarray} S_n(k_n) &=& -{8B k_n^6 \over 21} +{B^2 M k_n^{10} \over 27}
\bigg({6701 \over 350}+{27G-64\ln 2\over 11}\bigg)\nonumber \\ && 
+{B\beta_2 M k_n^6\over 315}(180G-191-48\ln2)+{4B\beta_4 Mk_n^8 \over 189}\Big(
8\ln2-7G- {259\over 30}\Big)\,.\end{eqnarray}
The expressions in the second lines of eqs.(67-70) stem from the interference 
terms of the Van-der-Waals interaction $V_{\rm vdw}$ with the (more long-range) 
$\beta_2$- and $\beta_4$-interactions. Finally, one notices that all 
interaction contributions derived here are in the form of even powers of the 
Fermi momentum, $k_f$ or $k_n$. It is expected that the analytical results 
compiled in this appendix could be useful for certain condensed matter systems.

\section*{Acknowledgements}
I thank G. Colo and J.W. Holt for informative discussions.

\end{document}